\documentclass[10pt,aps,prd,nofootinbib,showpacs,eqsecnum,superscriptaddress,longbibliography,preprintnumbers]{revtex4-2}
\usepackage{graphicx}
\usepackage{amsmath,amssymb}
\usepackage{amsfonts}
\usepackage{xspace} 
\usepackage[usenames]{color}
\usepackage{dcolumn}
\usepackage{bm}
\usepackage{mathrsfs}
\usepackage[colorlinks=true]{hyperref}
\usepackage[all]{hypcap} 
\usepackage[utf8]{inputenc} 
\usepackage{multirow}
\usepackage{etoolbox}
\usepackage{tikz}
\usepackage[normalem]{ulem}
\usepackage{cancel}

\newcommand{\dd}{\mathrm{d}}

\begin{document}

\title{An effective cosmological constant as black hole primary hair}

\author{Christos Charmousis}
\email{christos.charmousis@ijclab.in2p3.fr}
\affiliation{Université Paris-Saclay, CNRS/IN2P3, IJCLab, 91405 Orsay, France}

\author{Pedro G. S. Fernandes}
\email{fernandes@thphys.uni-heidelberg.de}
\affiliation{Institut f\"ur Theoretische Physik, Universit\"at Heidelberg, Philosophenweg 12, 69120 Heidelberg, Germany}

\author{Mokhtar Hassaine}
\email{hassaine@inst-mat.utalca.cl}
\affiliation{Instituto de Matemáticas, Universidad de Talca, Casilla 747, Talca, Chile}

\begin{abstract}
    We study Generalized Proca theories inspired by the recent regularised Proca theory of four-dimensional Gauss–Bonnet gravity. By abandoning the rigid constraints typically imposed by specific regularization schemes, we treat the coefficients of the terms in the action as free parameters. This approach uncovers a broader solution space that admits static and spherically symmetric black hole solutions characterized by primary hair, where, surprisingly, the cosmological constant arises naturally as a constant of integration even in the absence of a bare cosmological term.
\end{abstract}

\maketitle

\section{Introduction}

In theoretical physics, the construction of consistent and predictive models often involves introducing counterterms, additional terms in the action that cancel divergences arising in quantum field theories. These counterterms are not arbitrary; they are determined through symmetry principles, ensuring that the extended theory retains its fundamental properties. A notable example is the supersymmetrization procedure, where Noether's method is employed to derive the necessary counterterms, resulting in a theory where all coupling constants are fixed, providing a rigid and predictive framework \cite{Freedman:1976xh}.

Similarly, four dimensional Gauss-Bonnet regularization procedures, triggered from higher-dimensional Lovelock theories\footnote{Lovelock theories are the geometric generalization of General Relativity to higher dimensions featuring unique higher-order curvature scalars, giving second order equations of motion (see for example in this context \cite{Charmousis:2014mia}).}, have been developed recently, allowing a non-trivial contribution from the Gauss-Bonnet term in four dimensional spacetime. Indeed, the Gauss-Bonnet term, or more precisely, the generalised Euler-Chern density, is a key geometric scalar in Lovelock theory (see for example \cite{Deruelle:2003ck, Charmousis:2005ey}), giving trivial dynamics in four dimensional spacetime-as its geometric name hints. Gauss-Bonnet regularization procedures yield four dimensional scalar-tensor or vector tensor theories, where the coefficients of the terms in the action are given~\cite{Glavan:2019inb,Fernandes:2020nbq,Hennigar:2020lsl,Kobayashi:2020wqy,Lu:2020iav,Fernandes:2021dsb,Charmousis:2025jpx} (see Ref.~\cite{Fernandes:2022zrq} for a review). The values of the couplings are fixed, since deviations from the precise values of the coupling constants commonly result in the loss of underlying symmetries and hence integrability, or the breakdown of the theory's consistency~\cite{Fernandes:2021dsb}. Indeed, Einstein-Gauss-Bonnet (EGB) scalar tensor theories and their compact object solutions have been extensively studied in recent years \cite{Bakopoulos:2021liw, Charmousis:2021npl, Babichev:2022awg}. Nevertheless, few generalized theories to EGB along with their solutions have been found and are accompagnied by a number of limitations (consistent vacua, missing branches of solutions etc.) (see for example \cite{Babichev:2023dhs, Bakopoulos:2022csr, Bakopoulos:2025byy}). In this paper we dwell on this question but in regards to vector-tensor theories. 

In a previous work \cite{Charmousis:2025jpx}, we constructed a well-defined four-dimensional vector-tensor Proca Gauss-Bonnet (PGB) theory by applying a dimensional regularization procedure. The starting theory, was again, a Lovelock metric theory, however, endowed this time with a Weyl rather than a Levi Civita connection \cite{BeltranJimenez:2014iie, Bahamonde:2025qtc}.  The regularization procedure not only determines the allowed types of terms in the action but also fixes their relative coefficients. The resulting PGB theory contains three distinct terms with fixed coefficients, and is given by
\begin{eqnarray}
  \mathcal{L}_\mathcal{G}^{\rm VT} = 4 G^{\mu \nu}W_\mu W_\nu + 8 W^2 \nabla_\mu W^\mu + 6W^4,
  \label{ProcaReg}
\end{eqnarray}
where $W_\mu$ is the Proca field. Notice that although the theory stems from a higher dimensional Gauss-Bonnet term, the four dimensional Gauss-Bonnet scalar is absent from the resulting theory. Within this framework, we investigated static, spherically symmetric black hole solutions\footnote{Black hole solutions in generalised Proca theories have been addressed in several papers, see for example \cite{Chagoya:2016aar,Heisenberg:2017hwb,Heisenberg:2017xda, Babichev:2017rti}.}, and found that they carry primary hair~\cite{Charmousis:2025jpx}. Interestingly, the black hole solutions constructed in this framework have also been taken as a starting point for the study of gravitational wave echoes~\cite{Konoplya:2025uiq}. It has been shown that the PGB hair alters the perturbative structure of the spacetime in a way that supports late-time signal modulations. In this work, the echoes were not introduced through additional matter sources or modified boundary conditions, but rather emerged as a direct consequence of the intrinsic hairy geometry itself. By examining both the quasinormal spectrum and the time evolution of test fields, it was demonstrated that the presence of the primary Proca hair can by itself generate delayed repetitions in the ringdown signal, thus offering a new phenomenological window on this class of solutions.

The existence of black hole solutions with primary PGB hair has also motivated a series of follow-up studies exploring their broader implications. For instance, in Ref.~\cite{Lutfuoglu:2025ldc}, the classical and semiclassical properties of these spacetimes were investigated, with particular emphasis on horizon formation, geodesic motion, and phenomenological observables such as the shadow radius, Lyapunov exponent, and ISCO frequency, as well as the grey-body factors and quasinormal spectra. Extensions of the regularization procedure have also been considered, e.g. in Ref.~\cite{Eichhorn:2025pgy} where regular black holes were obtained, and in Ref.~\cite{Alkac:2025zzi} to lower-dimensional setups.
See also e.g. Refs.~\cite{Liu:2025dqg,Alkac:2025jhx,Lutfuoglu:2025qkt,Konoplya:2025bte,Fernandes:2025mic,Gurses:2026nvh} for other works inspired by the Proca theory of four-dimensional Gauss-Bonnet gravity.

Given the variety of interesting phenomena uncovered in the Proca Gauss-Bonnet theory derived in Ref.~\cite{Charmousis:2025jpx}, a natural question arises as to how strongly these features rely on the specific values of the coefficients determined by the dimensional regularization procedure. While the regularized theory is distinguished by its uniqueness and internal consistency, from a broader perspective in modified gravity it is instructive to relax this rigidity and treat the relative couplings between the different terms in Eq.~\eqref{ProcaReg} as free parameters. Doing so not only allows us to probe the robustness of the hairy black hole solutions, but also offers a way to explore potential phenomenological implications, since in realistic scenarios quantum corrections or embeddings in more fundamental frameworks could in principle alter the precise balance of coefficients.

In this work, we consider a generalized PGB theory with {\it{arbitrary relative couplings}} and analyze whether the distinctive features, most notably, the persistence of primary hair, survive in this enlarged parameter space. Unlike scalar-tensor EGB regularizations we find that this extension comes with no apparent difficulties nor any loss of integrability. 
Our approach provides a systematic means of disentangling which properties are intrinsic to the vector–tensor structure of the theory and which are consequence the specific coefficients fixed by the regularization scheme. Crucially in our study we will find that for more general couplings, an effective cosmological constant emerges as a pure, free integration constant allowing for de Sitter or anti-de Sitter asymptotics without any bulk cosmological constant. Our findings here may be quite relevant to cosmological dark energy and the old cosmological constant problem \cite{Weinberg:1988cp,Padilla:2015aaa} and the extension of self tuning ideas encountered in Fab 4 \cite{Charmousis:2011bf} for Horndenski theories \cite{Horndeski:1974wa}. We should emphasize that what we encounter here is quite different from self tuning scenarios for scalar tensor theories (see for example \cite{Appleby:2011aa, Charmousis:2015aya, Babichev:2016kdt, Babichev:2017lmw, Appleby:2018yci}) where an effective cosmological constant emerges from the collective geometric couplings of the modified gravity theory and is therefore highly constrained from strong coupling issues, phenomenological constraints, fine tuning problems etc. Since the cosmological constant is an integration constant here, phenomenological constraints would restrict its value locally in space and time rather than constraining the coupling constants of the theory in question. 

This paper is organized as follows. In Section~\ref{sectheory} we introduce the Proca Gauss-Bonnet-inspired theory with generalized couplings, and its static field equations in spherical symmetry. The field equations are solved in all generality, obtaining the unique  black-hole solutions with two primary hairs, one of which an effective cosmological constant. We also consider  the charged generalizations of these black holes by including the effects of the regularized four-dimensional Gauss-Bonnet scalar-tensor theory. In Section~\ref{sec:slow-rot} we generalize our results to slow-rotating solutions. We conclude in Section~\ref{sec:conclusions}.

\section{Proca Four-Dimensional theory and the general static and spherically symmetric solution}
\label{sectheory}

We consider a four-dimensional Proca theory, extending our previous study of the 4-dimensional Gauss-Bonnet regularization \cite{Charmousis:2025jpx}, which involves the Proca theory \eqref{ProcaReg} with arbitrary coefficients, namely
\begin{equation}
    S[g, W]  = \int d^4x \sqrt{-g} \Bigg[R   +\left(c_1 G^{\mu \nu}W_\mu W_\nu  +c_2 W^4+c_3 W^2 \nabla_\mu W^\mu \right) \Bigg].
    \label{eq:theory}
\end{equation}
For coupling constants $c
_1=4$, $c_2=6$ and $c_3=8$, the action reduces to the one studied in \cite{Charmousis:2025jpx}, and corresponding to the regularized action \eqref{ProcaReg}. We have used the shorthand notation $W^2=W_\mu W^\mu$ for the norm of the Proca field. The covariant field equations obtained from the variation of the above action are given in the appendix.

We now seek the general static and spherically symmetric solutions of \eqref{eq:theory}. Start with the general ansatz
\begin{equation}
\begin{aligned}
&\dd s^2 = -h(r) \dd t^2 + \frac{\dd r^2}{f(r)} + r^2\left( \dd \theta^2 + \sin^2\theta \dd \varphi^2 \right)\label{generic}, 
\qquad 
&W_\mu dx^\mu = w_{0}(r)\, \dd t+w_1(r)\,\dd r,
\end{aligned}
\end{equation}
where we note that the norm of the vector $W^2=W^2(r)$ is given by,
\begin{equation}
\label{norm}
    W^2=w_1^2 f - h^{-1} w_0^2.
\end{equation}
The general field equations for a static and spherically symmetric spacetime are also provided in the appendix. 
To simplify the field equations, we replace $w_0$ with $W^2$, using Eq. \eqref{norm}. In doing so the Proca equations become\footnote{We note that $E_{w_1}=0$ is equivalent to the metric equation $E_{tr}=0$.}
\begin{equation}
    2 c_1[r(f-1)]'+4 c_2 W^2 r^2+c_3\frac{(f h w_1^2 r^4)'}{r^2 w_1 h}=0 \label{int2},
\end{equation}
\begin{equation}
    2 c_1\left(\frac{f'}{f} - \frac{h'}{h} \right)+ c_3\frac{r  (W^2)'}{f w_1}=0,
\label{int1}
\end{equation}
where the prime denotes a radial derivative.
The second Proca equation \eqref{int1} gives us an important information: solutions are homogeneous ($f=h$) if and only if the norm of the Proca vector is $W^2=\mathrm{constant}$. Therefore for $f\neq h$ we must have $W^2$ non constant. Let us consider this possibility, whereupon $f\neq h$. Using \eqref{int1} to replace $h'$ in the remaining equations, three independent equations remain, which depend on $f$, $w_1$, $W^2$ and their first derivatives. Eq. \eqref{int2} becomes
\begin{equation}
\label{int3}
    4c_1\left[c_3 w_1 r^2 f+c_1 r (f-1) \right]'+r^2(c_3^2 (W^2)'r+8 c_1 c_2 W^2)=0,
\end{equation}
and a combination of $E_{tt}=0$, and $E_{rr}=0$, simplifies to give
\begin{equation}
\label{int4}
   4c_1\left[c_3 w_1 r^2 f+c_1 r (f-1) \right]'+8 c_1 c_2 r^2 W^2+(W^2)' r \left[r^2 c_3^2-\frac{4 f c_1^2 w_1 +r c_3 (c_1 W^2-2)}{\left(2 f w_1^2-W^2\right){w_1}} \right]=0.
\end{equation}

Taking the difference of \eqref{int3} and \eqref{int4} dictates that either $(W^2)'=0$ or  $4 c_1^2 f w_1+c_3 r (W^2 c_1-2)=0$. Therefore any non homogeneous solution must satisfy the second relation. Using this condition in \eqref{int4} gives $f'$ in terms of $W^2$ and inputting this in $E_{rr}=0$, implies that $W^2$ is constant. Therefore the only solutions to the equations of motion are homogeneous with $f=h$ and most importantly with a constant Proca field norm, $W^2=\mathrm{constant}$. Notice that \eqref{int3} can be integrated directly at the special point $\frac{8 c_1 c_2}{3 c_3^2}=1$ with $W^2$ not constant. This unfortunately does not yield any other non-trivial solutions. We will call henceforth this point, the \emph{Weyl point}.

Let us now without any loss of generality take $(W^2)'=0$, whereupon $f=h$. We now see from \eqref{int3} or \eqref{int4} that we get a first integral, resulting in
\begin{equation}
   4c_1(c_3 w_1 r^2 f+c_1 r (f-1))+\frac{8}{3}  c_1c_2 r^3 W^2=-8c_1^2(M-Q),
   \label{charge1}
\end{equation}
where the combination on the right-hand-side is an integration constant. This first integral corresponds to a symmetry of our set of equations of motion with $h(r)=f(r)$ and $W(r)=\mbox{cst}$ given by
\begin{equation}
    w_0^2\to w_0^2+\kappa(2rw_1 f+\kappa r^2+\frac{3c_3}{2c_2}f),\qquad w_1\to w_1+\frac{\kappa r}{f},\qquad W^2\to W^2-\frac{3c_3}{2c_2}\kappa,
\end{equation}
with $\kappa$ a free real parameter associated to the symmetry. This is a particular case of the symmetry found in \cite{Charmousis:2025jpx}. The relation \eqref{charge1} defines a conserved charge and a symmetry which is present whenever spacetime is homogeneous, $f=h$.

Solving Eq.~\eqref{charge1} for $w_1$, we obtain
\begin{equation}
    w_1=\frac{c_1}{c_3 r f}\left(1-f+\frac{2(Q-M)}{r}-\frac{2c_2}{3c_1}\lambda r^2 \right)\label{solw1},
\end{equation}
which when used in the last remaining independent equation gives us a first order non linear ODE with respect to $f$ which can be integrated once to give the general solution
\begin{equation}
\label{solfM}
     f(r)= 1-\frac{2(M-Q)}{r}+\frac{r^2}{2\alpha}\left[1-\frac{\lambda}{2}\beta\pm\sqrt{1-\lambda\beta  \left(1-\frac{c_1\lambda}{4}\right)+\frac{8\alpha}{r^3}\left(Q+\frac{\lambda(M-Q)}{2}\beta\right)}\right],
\end{equation}
where we have defined
\begin{equation}
    \lambda = W^2, \qquad \beta = \left( 1 - \frac{8c_1 c_2}{3c_3^2} \right)c_1, \qquad \alpha = -\frac{c_1^3}{c_3^2},
\end{equation}
to write the metric function in a compact form. Note that while $\beta$ and $\alpha$ are fixed by the theory, $\lambda$ is a free parameter as it is an integration constant. The parameters $M$ and $Q$ are also free integration constants. It is remarkable that for generic couplings $c_1$, $c_2$ and $c_3$ we were able to integrate the field equation exactly. This is unlike, for example, the case of the scalar-tensor version of four-dimensional Gauss-Bonnet theories, where any deviation from the couplings obtained through dimensional regularizations results in non-integrability of the field equations~\cite{Fernandes:2021dsb}.

It is now evident that the solution found in \cite{Charmousis:2025jpx} corresponds to a theory  with $\beta=0$. Whenever $\lambda=Q=M=0$ we have a flat spacetime metric for the minus branch, c.f. the $\pm$ factor in Eq.~\eqref{solfM}, which we therefore choose to keep from now on. We can also write the above metric in a Boulware-Deser form~\cite{Boulware:1985wk}. To do this we define $\hat \alpha=\frac{\alpha}{1-\frac{\lambda\beta}{2}}$ and we get,
\begin{eqnarray}
\label{BD}
     f(r)= 1-\frac{2(M-Q)}{r}+\frac{r^2}{2\hat \alpha}\left[1\pm\sqrt{1+\frac{\beta \lambda^2({c_1}-\beta)}{(2-\beta \lambda)^2}+\frac{8\hat \alpha}{r^3}\left(Q+M\frac{\lambda\beta}{2-\lambda \beta}\right)}\right].
\end{eqnarray}

When $\beta=0$, the parameter $M$ represents the ADM mass of the solution, while $Q$ is a Proca charge akin to the one found in Ref.~\cite{Charmousis:2025jpx}. However, when $\beta\neq 0$, the role of all charges is intertwined. What stands out in the general case, is the presence of an additional integration constant, $\lambda\equiv W^2$, which assumes the role of an effective cosmological constant (for theories with $\beta \neq 0$). Indeed to see this, we momentarily switch off $M=Q=0$, to get a static locally maximally symmetric metric with an effective cosmological constant of magnitude
\begin{equation}
    \Lambda_{{\tiny \mbox{eff}}}=\frac{3}{2\alpha}\left(\sqrt{1-\lambda \beta \left(1-\frac{{{c_1}}\lambda}{4} \right)}-1+\frac{\lambda \beta}{2} \right),
\end{equation}
Again we have kept "minus" branch in order to ensure a Minkowski vacuum for $\lambda\to 0$. 

The interesting outcome of the above discussion is that theories with $\beta\neq 0$ produce an effective cosmological constant. It is important to emphasize that this cosmological constant is an integration constant and does not originate from a bulk cosmological constant term. It is in fact trivial to include a bulk cosmological constant term in the action and reproduce the solution. Note that on shell the Proca term  $c_2 W^4$ in the action plays the role of an effective cosmological constant as we saw that dynamically the norm of the solution is always constant.  The only case, apart from $\beta=0$, escaping this rule is the theory with $\beta=c_1$ whereupon $c_2=0$ and we have again an asymptotically flat solution. In fact this asymptotically flat solution is almost identical to that appearing in \cite{Charmousis:2025jpx} upon a rescaling of the ADM mass by a factor dependent on $\lambda$. In this case, we get
\begin{equation}
      f(r)= 1-\frac{2(M-Q)}{r}+\frac{r^2}{2\hat\alpha}\left[1\pm\sqrt{1+\frac{8\hat\alpha}{r^3}\left(Q+\frac{{c_1}\lambda M}{2-{c_1}\lambda}\right)}\right],
\end{equation}
where now we have $\hat \alpha=\frac{2\alpha}{2-c_1\lambda}$.
When we take $M=Q$ we get an EGB type of black hole with an effective  cosmological constant.

When $c_1$ or $c_3$ are zero (but not both) we have a de Sitter Schwarschild black hole spacetime where again the role of the cosmological constant is assumed by the $c_2$ term in the action. Specifically for $c_1=0$ the general solution reads, 
\begin{equation}
    f=h=1-\frac{2M}{r}+c_2 \frac{\lambda^2}{6}r^2,\qquad w_1=\frac{Q-\frac{4 r^3 \lambda c_2}{c_3}}{6r^2 f(r)}.
\end{equation}

In turn, for $c_3=0$, we obtain a slightly different configuration that reads,
\begin{equation}
    f=h=1-\frac{2M}{r}-\frac{2c_2\lambda}{3c_1}r^2,\qquad w_1=\pm \frac{\sqrt{Qr-\frac{3c_2\lambda}{2}(\lambda c_1-4)r^4}}{3rc_1 f(r)}.
\end{equation}
In each case we observe that the Proca charge $Q$ is stealth for the metric (i.e., only shows up in the profile for the Proca field), while the cosmological constant is not stealth which is exactly the opposite of what happens for the case $\beta=0$.

\subsection{Charged black holes with primary hair, and the scalar-tensor EGB theory}

For completeness, we also present the charged generalizations of the black holes discussed in previous section by including the effects of the regularized four-dimensional Gauss-Bonnet scalar-tensor theory, and a bare cosmological constant term $\bar{\Lambda}$. To this end, we consider the action
\begin{equation}
\label{genProca}
S[g, W, A,\phi] = \int d^4x \sqrt{-g} \left[ R-2\bar{\Lambda} + \left( c_1\, G^{\mu\nu} W_\mu W_\nu 
+ c_2 W^4 + c_3 W^2 \nabla_\mu W^\mu \right)+\alpha_1\,F_{\mu\nu}F^{\mu\nu}+\alpha_2\,\mathcal{L}_{G}^{ST} \right],
\end{equation}
where
\begin{equation}
\begin{aligned}
    \mathcal{L}_{\mathcal{G}}^{\mathrm{ST}}=&\phi \mathcal{G}-4 G^{\mu \nu} \nabla_\mu \phi \nabla_\nu \phi-4 \square \phi(\partial \phi)^2-2(\partial \phi)^4,\\
    F_{\mu\nu}=&\partial_\mu\,A_{\nu}-\partial_\nu\,A_{\mu},
\end{aligned}
\end{equation}
We consider the following ansatz,
\begin{equation}
\begin{aligned}
ds^2 =& -f(r)dt^2 + \frac{dr^2}{f(r)} + r^2 d\Omega_2^2, 
\qquad 
&W_\mu dx^\mu = w_0(r) dt + w_1(r) dr,\\
    \phi=&\phi(r), &A_\mu \mathrm{d} x^\mu=a_0(r) \mathrm{d} t,
\end{aligned}
\end{equation}
and from the $\phi$ and $A_\mu$ equations of motion, we find that the scalar and gauge fields are given by
\begin{equation}
    \phi^{\prime}=\frac{1-\sqrt{f}}{r \sqrt{f}},\qquad a_0=\frac{Q_e}{r}+\mathrm{constant},
\end{equation}
while the metric function for $\alpha\neq \alpha_2$ reads
\begin{equation}
    f(r) = 1
- \frac{2 (M - Q)\,\alpha}{r\,(\alpha - \alpha_2)}
+ \frac{r^{2}}{2(\alpha - \alpha_2)}
\Bigg(
1
- \frac{\beta \lambda}{2}
\pm \sqrt{
1
- \lambda \beta \left(1 - \frac{c_1 \lambda}{4}\right)
- \frac{1}{4} (c_1 - \beta)
\frac{\alpha_2\lambda^{2} \beta}{\alpha}
+ \frac{4}{3}\bar{\Lambda}(\alpha - \alpha_2)
+ P(r)
}
\Bigg),
\end{equation}
where we have defined 
\[
P(r)=\frac{8 {\alpha}(M-Q)(\frac{\lambda\beta}{2}-1)+({\alpha}-\alpha_2)\frac{8M}{{\alpha}}}{r^3}+\frac{Q_e^2\alpha_1}{r^4}({\alpha}-\alpha_2)+\frac{16(M-Q)^2 {\alpha} \alpha_2}{r^6}.
\]
Note that the expression for the Proca field remains unchanged as in the previous case.

On the other hand, for $\alpha= \alpha_2$, the quadratic equation determining $f$ becomes linear whose solution becomes 
\begin{equation}
    f(r) =
1
+ \frac{
- \dfrac{4 \alpha (M - Q)^{2}}{r}
+ \dfrac{Q_{e}^{2} \alpha_{1} r}{4} 
+ \left(\dfrac{\Lambda}{3} 
+ \left(\dfrac{\lambda }{4}\right)^2\frac{\beta (c_{1} - \beta)}{\alpha} \right)r^5
+ 2 M r^{2}
}{
4 (M - Q) \alpha
- r^{3} \left(1 - \dfrac{\beta \lambda}{2}\right)
}.
\end{equation}

\section{Slowly-rotating black hole}
\label{sec:slow-rot}
Having obtained exact static solutions, it is quite natural to investigate whether slowly rotating configurations exist, especially since stationary axisymmetric solutions with $W^2=0$ have recently been found in these theories~\cite{Fernandes:2026rjs}. This complements previous results for the model defined by $c_1=0=c_2$ in \eqref{eq:theory} where stealth Kerr solutions were found for which the norm of the Proca is also zero \cite{Cisterna:2016nwq}. Also very recently slowly rotating solutions have been found in a wide range of scalar tensor theories \cite{Candan:2025fbl}. Following the standard approach for slow-rotation, we consider $g_{t\varphi}-$component that is linear in the angular momentum parameter $a$, assumed to be small compared to the mass of the black hole. Simultaneously, we consider a component of the vector field $W_\mu$ in the $\varphi$-direction, as is the case, for example, for the slowly rotating limit of the Kerr-Newman solution. Drawing inspiration from the static case, we require the norm of the vector field to be constant while working strictly at the first order in $a$. Under these conditions, we consider the following ansatz for the slowly-rotating metric 
\begin{equation}
    ds^2 = -f(r) dt^2 + \frac{dr^2}{f(r)} + r^2\left( d\theta^2 + \sin^2\theta d\varphi^2 \right)+a h_1(r)\sin^2\theta\,dtd\varphi,
    \label{eq:metric}
\end{equation}
with a Proca field given by
\begin{equation}
    W_\mu \dd x^\mu = w_0(r) \dd t + w_1(r) \dd r+a w_2(r)\sin^2\theta d\varphi.
    \label{eq:ansatzVF}
\end{equation}
The condition of constant norm of the vector field $W^2\equiv \lambda$ at first order in $a$ becomes
\begin{eqnarray}
\label{nullcond}
   f w_1^2-\frac{w_0^2}{f}=\lambda. 
\end{eqnarray}

We now proceed to the derivation of the field equations, starting with the vector field equation $E_{W^{\varphi}}$. At first order in $a$, this equation reduces to
\begin{equation}
\begin{aligned}
    E_{W^{\varphi}}=&\frac{ac_1}{2r^4f}\left[2h_1w_0(r f'-1)+2w_2f(r^2 f')'+r^2f w_0 h_1''\right]\\
    &+\frac{2ac_2}{r^2f^2}\left[(f\lambda)(h_1w_0+2fw_2)\right]
    +\frac{ac_3}{r^3f}\left[\frac{(r^2fw_1)'}{r}(h_1w_0+2fw_2)\right],\label{primaJ4}
\end{aligned}
\end{equation}
while the $ E_{t\varphi}$ Einstein equation at first order reads
 \begin{equation}
\begin{aligned}
     E_{t\varphi}=&\frac{a\sin^2\theta}{4r^2}\left[rh_1(fr)''-f(r^2h_1''-2h_1)\right]-\frac{a\,c_2}{4\,{f^{2}}}\sin^2\theta\left[(f\lambda)(h_1(f\lambda)-8f\,w_0w_2)\right]\\
     &+\frac{a\,c_3}{4\,r\,f}\sin^2\theta\left[\frac{4\,fw_0w_2}{r}(r^2fw_1)'-r\,h_1w_1f^2\left(\lambda\right)'\right]+\frac{a\,c_1}{8}\,\sin^2\theta\bigg\{h_1(f\,\lambda)\left(\frac{f''}{f}-\frac{h_1''}{h_1}\right)
     -h_1' f\left(f\,\lambda\right)'\\
     &+4\,f(w_0w_2)''+\frac{8}{r^2}w_0w_2(r f'-1)-h_1\bigg[\frac{r}{f^2}\left(\frac{f^3\,\lambda}{r^2}\right)'-f'\left(\lambda\right)'{-}4\sqrt{r}\left(\frac{(w_0^2)'}{2\,\sqrt{r}}\right)'{-}\frac{10\,f\,w_1}{r}(fw_1)'\bigg ]\bigg\}.\label{primaE14}
\end{aligned}
\end{equation}
From the Eq. \eqref{primaJ4}, we can obtain an expression for $w_2$ in terms of the constant norm $\lambda$
{
\begin{equation}
   w_2= -\frac{
\sqrt{
9 c_1^2 \left( 2 (Q - M) + r - r f \right)^2
+ 4 c_2^2 r^6 \lambda^2
+ 3 r^3 \lambda \left(
8 c_1 c_2 (M - Q)
- 4 c_1 c_2 r
+ \left(4 c_1 c_2 - 3 c_3^2 \right) r f
\right)
}
\left(
-2 h_1 + r^2 h_1''
\right)
}{
6 c_3 r^2 \left(
2 - 2 f + r^2 f''
\right).\label{w2lambda1}
}
\end{equation}
}
and, using the expression previously obtained for $w_0$, this can be more conveniently re-written as
\begin{equation}
w_2=-\frac{w_0}{2}\frac{\left(
  r^2 h_1''-2 h_1
\right)}{\left(
2 - 2 f + r^2 f''\right)}=-\frac{w_0}{2}\frac{[r^4(\frac{h_1}{r^2})']'}{[r^4(\frac{1-f}{r^2})^{'}]'},
\end{equation}
which has the global symmetry $h_1\rightarrow h_1+c\,r^2$.
Substituting the expression for $w_2$ into the Einstein equation \eqref{primaE14} generally leads to a fourth-order differential equation for $h_1$. However, the order of this system can be reduced to two by imposing the following condition
\begin{equation}
    2h_1 - r^2 h_1'' = A \left( 2 - 2f + r^2 f'' \right), \label{eqh}
\end{equation}
where $A$ is a constant determined by the specific form of $w_2$. This assumption is physically motivated by the results in \cite{Fernandes:2026rjs}, where the vector field is shown to be proportional to the null congruence
\begin{equation}
    l_\mu \dd x^\mu = \dd t - a \sin^2\theta \, \dd \varphi + \frac{r^2+a^2\cos^2\theta}{r^2+a^2} \, \dd r \sim \dd t+\dd r - a \sin^2\theta \, \dd \varphi + \mathcal{O}(a^2).
\end{equation}
This identification implies that the radial  $w_0(r)$ and angular $w_2(r)$ components, as defined in our slow-rotation expansion \eqref{eq:ansatzVF}, must be proportional. Hence, assuming \eqref{eqh}, the $E_{t\varphi}$ component of the Einstein equations reduces to
\begin{equation}
\label{etp}
    E_{t\varphi} = -\frac{a A \sin^2\theta}{2r^2} \lambda c_1 f,
\end{equation}
where $f$ is the static seed solution \eqref{solfM}. Excluding the trivial case $c_1=0$, the field equation $E_{t\varphi}=0$ is satisfied if either the norm of the vector field vanishes ($W^2=0$, or $\lambda=0$) or the vector perturbation vanishes ($w_2=0$), which occurs when $A=0$.

Let us first consider the case where $\lambda=0$. From \eqref{w2lambda1} and the assumption  \eqref{eqh}, one obtains that 
\begin{equation}
   w_2= \pm A\frac{
 c_1 \left( 2 (Q - M) + r - r f \right)
}{
2 c_3 r^2. \label{w2lambda}
}
\end{equation}
Hence, the full system is satisfied for the metric function $f$ given by \eqref{solfM} with $\lambda=0$. 

On the other hand, the function $h_1(r)$ given by (\ref{eqh}) can be integrated as 
\begin{equation}
    h_1(r) = \mp \frac{A}{2\alpha} \sqrt{r^4 + 8\alpha Q r} + \frac{C_4}{r} + r^2 C_5,
\end{equation}
where $C_4$ and $C_5$ are constants of integration, and, as usual the constant $C_5$ can be gauged out. The auxiliary function $w_2$ is given by
\begin{equation}
    w_2 = \pm \frac{A c_1}{4\alpha r c_3} \left( \sqrt{r^4 + 8\alpha Q r} - r^2 \right).
\end{equation}

The slowly rotating configurations obtained here represent the slow-rotation limit of the stationary axisymmetric solutions previously derived in \cite{Fernandes:2026rjs}, when $W^2\equiv \lambda=0$. A crucial distinction arises: while the full stationary solutions in \cite{Fernandes:2026rjs} contain an arbitrary function of the angular coordinate $\theta$, our derivation naturally constrains this dependency, leaving no such freedom. This is a consequence of assuming spherical symmetry in the slowly-rotating limit.

Alternatively, the Einstein equation \eqref{etp} can also be satisfied by setting $A=0$, which implies $w_2=0$. In this instance, the metric function $h_1$ reduces to
\begin{equation}
    h_1 = \frac{\hat{C}_1}{r}+\hat{C}_2 r^2
\end{equation}
By eliminating, as usual the  quadratic correction $\hat{C}_2=0$, we recover the standard Lense-Thirring effect without introducing a  vector perturbation. This indicates that the static solution, despite its non-vanishing vector field norm, admits a slowly rotating extension whose $g_{t\varphi}$ component is identical to that of General Relativity. This result points toward the potential existence of a broader class of rotating solutions generalizing those in \cite{Fernandes:2026rjs}, characterized by a non-zero but constant vector field norm.

To conclude this section, we consider the Proca model augmented with the Maxwell term, corresponding to Eq. \eqref{genProca} with $\alpha_2=0$. By proceeding in a similar manner, one finds that the $E_{t\varphi}$ field equation becomes 
\begin{equation}
\label{etp2}
    E_{t\varphi} = \frac{a A \sin^2\theta}{2r^2} \left(\frac{Q_e^2\alpha_1}{2r^2}-\lambda c_1\right) f.
\end{equation}
Consequently, for a non-vanishing proportionality constant $A \neq 0$, both the norm of the vector field $\lambda$ and the electric charge $Q_e$ must vanish. In this limit, the system effectively reduces to the previously discussed case. In analogy with the previous case, the $A=0$ configuration describes a static solution that, despite possessing a non-vanishing vector field norm and charge, permits a slowly rotating extension. This suggests the potential existence of fully rotating configurations within our Proca theory when supplemented by a Maxwell term, that is for  Eq. \eqref{genProca} with $\alpha_2=0$.

\section{Conclusions}
\label{sec:conclusions}
In this article, we have systematically explored and characterized a novel class of black hole solutions within the framework of PGB theories. By relaxing the constraints inherited from the regularization procedure, we have considered the coupling constants of the theory as generalized free parameters. We have shown that, despite the departure from the specific ``tuned" structure of regularized Gauss-Bonnet gravity,  static and spherically symmetric black hole solutions with primary hair endure as a fundamental feature of Proca theories of this kind.

One of the results in our work is the emergence of an effective cosmological constant, which crucially appears as a free integration constant, even in the absence of an explicit bare cosmological term. This mathematical feature carries physical implications regarding the cosmological constant problem \cite{Weinberg:1988cp,Padilla:2015aaa,Charmousis:2011bf}. The PGB theory dynamically accommodates asymptotically de-Sitter of anti de-Sitter spacetime solutions without the need for a bulk cosmological constant. Such a mechanism is not entirely without precedent. Indeed, in \cite{Babichev:2017rti}, it was shown that for the standard Proca theory supplemented by the $c_1-$term of \eqref{eq:theory}, the Proca mass can play the role of an effective cosmological constant, distinctly different from a bare cosmological constant. Our current results, however, suggest a more fundamental ability of the  gravitational field to self-generate its own asymptotic background pointing toward a deeper connection between the Proca sector and the underlying spacetime curvature, a relationship that deserves further investigation.  

Black holes with primary hair have been also found \cite{Bakopoulos:2023fmv, Baake:2023zsq, Bakopoulos:2023sdm, Charmousis:2025xug} in higher order scalar tensor theories which crucially possess shift symmetry for the scalar field. This global symmetry (for a general discussion see \cite{Babichev:2024krm}) enables the presence of an independent integration constant which in turn spontaneously breaks local Lorentz invariance as we have encountered here for the Proca field. Although the mechanisms are quite different it would be interesting to study the similarities, consequences and constraints for the primary hair in question originating from spontaneous Lorentz symmetry breaking.

We have also demonstrated the versatility of the framework by extending the class of solutions by incorporating electromagnetic charges and adding scalar–tensor counterparts of the regularized Gauss–Bonnet framework. It is worth stressing that, usually, once one departs from the specific structure inherited from the regularization procedure, the construction of explicit solutions typically becomes in general highly nontrivial or even impossible. This was observed, for example, in the scalar-tensor versions of four-dimensional Gauss-Bonnet gravity, see e.g.~\cite{Fernandes:2021dsb}. Altogether, our results highlight  a surprising persistence of integrability within the Proca sector, ensuring that black hole solutions with primary hair remain robust even when additional sources are incorporated.

An intriguing aspect that deserves further investigation is that all static and spherically symmetric solutions feature a constant norm for the Proca vector field. From a mathematical perspective, the constancy of the norm considerably simplifies the equations of motion, facilitating the construction of explicit solutions. It may also point to an implicit or hidden symmetry of the theory. From a physical point of view, a Proca field with constant norm could correspond to a stable configuration that minimizes the energy associated with the vector field. Moreover, a constant norm for the Proca field spontaneously breaks Lorentz invariance, as the Proca field defines a preferred direction in spacetime.
This raises questions regarding the interpretation of the primary hair: does this constancy of $W^2$ reflect a restriction on the type of primary charges that black holes can carry in these theories? How does this constraint affect the thermodynamic properties of the solutions, such as their stability or response to perturbations? Exploring this aspect could shed light on the fundamental mechanisms that allow for the existence of primary-hair solutions in vector–tensor theories. Future works could explore the stability of these solutions, and possible observational signatures, particularly in the context of gravitational wave physics and black hole shadow phenomenology \cite{Konoplya:2025uiq}. Most probably the more interesting aspects coming forth from this work are the implications to the cosmological constant in relation to dark energy scenarios and the possible phenomenological constraints on its value at different epochs of our Universe.

\section*{Acknowledgments}
We are very happy to acknowledge interesting and enlightening discussions with Eugeny Babichev, Jacopo Mazza and Tony Padilla.
C.C. acknowledges partial support by the French National Research Agency (ANR) via Grant No.
ANR-22-CE31-0015-01 associated with the project Strong. P.~F. acknowledges funding by the Deutsche Forschungsgemeinschaft (DFG, German Research Foundation) under Germany’s Excellence Strategy EXC 2181/1 - 390900948 (the Heidelberg STRUCTURES Excellence Cluster). The work of M.~H. is partially supported by
FONDECYT grant 1260479.

\appendix
\section{Field equations}
In this appendix, we present the Einstein and vector field equations. We first provide their general covariant form, followed by their specific expressions under a spherically symmetric ansatz. For a generic action in $4$ dimensions given by
\begin{equation}
    S = \int d^4x \sqrt{-g} \Bigg[R   +\left(c_1 G^{\mu \nu}W_\mu W_\nu  +c_2 W^4+c_3 W^2 \nabla_\mu W^\mu \right) \Bigg],
\end{equation}
the field equations are given by $G_{\mu\nu}=T_{\mu\nu}$ where the stress energy-momentum tensor reads
\begin{equation}
\begin{aligned}
     T_{\mu \nu}=&c_1 \Big[R_{\mu\rho\nu\sigma}W^\rho W^\sigma+(\nabla_\nu \nabla_\mu W_{\rho})W^\rho+\frac{1}{2}\Big( R_{\nu\rho}W_\mu W^\rho-R_{\mu \nu}\,W^2+R_{\mu \rho}W^\rho W_\nu-R \,W_\mu W_\nu\\
     &+(\square W_\mu)W_\nu+W_\mu (\square W_\nu)+2(\nabla_\rho W_\mu)(\nabla^\rho W_{\nu})-(\nabla_\mu W_\rho)(\nabla^\rho W_\nu)-(\nabla_\mu W_\nu)(\nabla_\rho W^\rho)\\
     &-(\nabla_\mu\nabla_\rho W^\rho)W_\nu-(\nabla_\mu\nabla_\rho W_\nu)W^\rho-(\nabla^\rho W_\mu)(\nabla_\nu W_\rho)+2(\nabla_\mu W^\rho)(\nabla_\nu W_\rho)-(\nabla_\nu W_\mu)(\nabla_\rho W^\rho)\\
     &-W_\mu(\nabla_\nu \nabla_\rho W^\rho)-(\nabla_\nu \nabla_\rho W_\mu)W^{\rho}\Big)+g_{\mu \nu}\big(\frac{1}{4}R\,W^2-R_{\rho \sigma}W^{\rho}W^{\sigma}+W^{\rho}(\nabla_\sigma \nabla_\rho W^{\sigma})
     \\&+\frac{1}{2}(\nabla_\rho W^{\rho})^2-W^{\rho}(\square W_{\rho})+\frac{1}{2}(\nabla_\rho W_\sigma)(\nabla^{\sigma}W^{\rho})-(\nabla_\rho W_\sigma)^2\big)\Big]\\
     &+c_3\left[W_\mu W_\nu(\nabla_\rho W^\rho)-(\nabla_\mu W_\rho)W_\nu W^\rho-W_\mu(\nabla_\nu W_\rho)W^\rho+g_{\mu \nu}W^\rho W^\sigma(\nabla_\sigma W_\rho)\right]\\
     &+c_2\left[2W^2W_\mu W_ \nu-\frac{1}{2}g_{\mu \nu}W^4\right].
\end{aligned}
\end{equation}
The vector field equation is 
\begin{equation}
\begin{aligned}
2\,c_2\,  W^{2} W^{\mu} +c_1\, W_{\alpha} G^{\mu\alpha} 
+ \,c_3\left( W^{\mu} \nabla_{\alpha} W^{\alpha} 
-  W_{\alpha} \nabla^{\mu} W^{\alpha}\right) &= 0. 
\end{aligned}
\label{vfeq}
\end{equation}

For a general spherical ansatz of the form 
\begin{eqnarray}
    ds^2=-h(r)dt^2+\frac{dr^2}{f(r)}+r^2\left(d\theta^2+\sin^2\theta d\varphi^2\right), \qquad W_\mu dx^\mu = w_{0}(r)\, dt+w_1(r)\,dr
\end{eqnarray}
the field equations are given by

{\begin{equation}
\begin{aligned}
E_{tt} ={}&
-\frac{c_2}{2 r^2 h}
\left(3 r w_0^2 + r f h w_1^2\right)
\left(r w_0^2 - r f h w_1^2\right) \\[6pt]
&+ \frac{c_3}{2}
\Bigg[
 f h w_1 
\left(
\frac{1}{h}\left( w_0^2 - f h w_1^2 \right)
\right)'
+ 2\,\frac{w_0^2}{r^2}(r^2 f w_1)'
+ \frac{w_0^2 w_1 f^2}{h}
\left(\frac{h}{f}\right)'
\Bigg] \\[6pt]
&+ \frac{c_1}{2 r^2}
\Big[
w_0^2\left((r f)'-1\right)
- f h (r f w_1^2)'
- h r (f^2 w_1^2)'
- f h w_1^2
\Big]
- \frac{h}{r^2}\left((r f)'-1\right),\\
\nonumber\\
E_{tr}={}&
w_0 w_1
\Bigg[\frac{c_1}{r^2}
\left(
\frac{f}{h}(r h)' - 1
\right)
+ \frac{2 c_2}{h}\left(f h w_1^2 - w_0^2\right) 
+ \frac{c_3}{2 w_1}
\left(
\frac{w_1^2}{r^4}\,\frac{f}{h}(r^4 h)'
+ \left(\frac{w_0^2}{h}\right)'
\right)
\Bigg],\\
\nonumber\\
E_{rr}={}&
-\frac{c_2}{2 f h^2}
\left(w_0^2 + 3 f h w_1^2\right)
\left(w_0^2 - f h w_1^2\right)
+ \frac{c_3}{2}\, w_1
\left[
\left(\frac{w_0^2}{h}\right)'
+ \frac{f}{h}\,\frac{w_1^2}{r^4}(r^4 h)'
\right] \\[6pt]
&+ \frac{c_1}{2 r^2 f h}
\Big[ 2 r f\, (w_0^2)'
- w_0^2\!\left(1 + \frac{f}{h}\, r^2 \left(\frac{h}{r}\right)'\right)
+ f^2 w_1^2\!\left(3 (r h)' - \frac{h}{f}\right)
\Big]
+ \frac{(r h)' - \tfrac{h}{f}}{r^2 h},\\
\nonumber\\
E_{\theta\theta}={}&
- \frac{c_2 r^2}{2 h^2}
\left( w_0^2 - f h w_1^2 \right)^2
- \frac{c_3 r^2 f w_1}{2}
\left[
\left(\frac{w_0^2}{h}\right)'
- (f w_1^2)'
\right] \\[6pt]
&- c_1
\Bigg[\frac{r}{2}
\left(
\frac{f}{h}\left(3 w_0^2 - f h w_1^2\right)
\right)'
- \frac{f}{2h}(r^2 w_0^2)'' \\[6pt]
&\qquad
+ \frac{r^2 f}{4 h^2}
\left(
h' \left( w_0^2 - f h w_1^2\right)
\right)'
- \frac{r^2 f}{4 h^3}
(h')^2
\left( w_0^2 - f h w_1^2\right) \\[6pt]
&\qquad
+ \frac{f}{h}\, w_0^2
- \frac{1}{4}\left(\frac{f}{h}\right)'
\Bigg(
7 r w_0^2 - r f h w_1^2
- \frac{r^2 w_0^2}{2 h}\,h'
+ \frac{r^2 f w_1^2}{2}\,h'
+ 2 r^2 w_0 w_0'
\Bigg)
\Bigg]\\
&+\frac{1}{2}\left(\frac{f}{h}r^2 h'\right)'
-\frac{r^4}{4}\left(\frac{h}{r^2}\right)'\left(\frac{f}{h}\right)'.
\end{aligned}
\end{equation}}
On the  other side, the non-zero components of the equation of motion for the vector field $W$ read
{\begin{equation}
    \begin{aligned}
        \epsilon_{Wt}= {}&
- \frac{2 w_0}{h}
\Bigg[
\frac{c_1}{r^2}\left((r f)' - 1\right)
+ \frac{2 c_2}{h}\left(f h w_1^2 - w_0^2\right) 
+ c_3\left(
\frac{1}{r^2}(r^2 f w_1)'
- \frac{w_1 h}{2}\left(\frac{f}{h}\right)'
\right)
\Bigg],\\
        \epsilon_{Wr}={}&
\frac{2 c_1 f^2 w_1}{r^2 h}
\left((r h)' - \frac{h}{f}\right)
+ \frac{4 c_2 f w_1}{h}
\left(f h w_1^2 - w_0^2\right)+ c_3
\left[
\left(\frac{f}{h}\right)^2
\frac{w_1^2}{2 r^8}(r^8 h^2)'
+ f\left(\frac{w_0^2}{h}\right)'
\right].
    \end{aligned}
\end{equation}}

\bibliography{References}

\end{document}